\documentclass[amsmath,amssymb,pra,twocolumn,groupaddress]{revtex4-2}
\usepackage{graphicx}
\usepackage{dcolumn}
\usepackage{bm}
\usepackage[colorlinks,linkcolor=blue,anchorcolor=blue,citecolor=blue,]{hyperref}
\usepackage{xcolor}

\begin{document}

\title{Precision Measurement of Vibrational Quanta in Tritium Hydride (HT)}

\author{F. M. J. Cozijn}
\affiliation{Department of Physics and Astronomy, LaserLab, Vrije Universiteit\\
De Boelelaan 1081, 1081 HV Amsterdam, The Netherlands}

\author{M. L. Diouf}
\affiliation{Department of Physics and Astronomy, LaserLab, Vrije Universiteit\\
De Boelelaan 1081, 1081 HV Amsterdam, The Netherlands}

\author{W. Ubachs}
\affiliation{Department of Physics and Astronomy, LaserLab, Vrije Universiteit\\
De Boelelaan 1081, 1081 HV Amsterdam, The Netherlands}
 \email{w.m.g.ubachs@vu.nl}

\author{V. Hermann}
\affiliation{Tritium Laboratory Karlsruhe, Institute of Astroparticle Physics, Karlsruhe Institute of Technology, Hermann-von-Helmholtz-Platz 1,
76344 Eggenstein-Leopoldshafen, Germany}

\author{M. Schlösser}
\affiliation{Tritium Laboratory Karlsruhe, Institute of Astroparticle Physics, Karlsruhe Institute of Technology, Hermann-von-Helmholtz-Platz 1,
76344 Eggenstein-Leopoldshafen, Germany}

\date{\today}

\begin{abstract}
\noindent
Saturated absorption measurements of transitions in the (2-0) band of radioactive tritium hydride (HT) are performed with the ultra-sensitive NICE-OHMS intracavity absorption technique in the range 1460-1510 nm. 
The hyperfine structure of rovibrational transitions of HT, in contrast to that of HD, exhibits a single isolated hyperfine component, allowing for the accurate determination of hyperfineless rovibrational transition frequencies, resulting in R(0) = $203\,396\,426\,692$ (22) kHz and R(1) = $205\,380 \,033 \,644$ (21) kHz. 
This corresponds to an accuracy three orders of magnitude better than previous measurements in tritiated hydrogen molecules. 
Observation of an isolated component in P(1) with reversed signal amplitude contradicts models for line shapes in HD based on cross-over resonances.

\end{abstract}

\maketitle

In the recent decade rapid progress has been made in precision investigations of the quantum level structure of the smallest neutral molecular entity, the hydrogen molecule, existing as three stable isotopologues. 
On the theory side non-relativistic quantum electrodynamics (QED) 
has been developed into the framework of nonadiabatic perturbation theory (NAPT)~\cite{Czachorowski2018,Komasa2019}. This has resulted in a publicly available  code for computing level energies in molecular hydrogen~\cite{SPECTRE2022}.
In parallel, so-called pre-BO methods have been developed that treat the 4-particle system in a direct variational approach~\cite{Simmen2013,Wang2018} leading to the most accurate binding energies of molecular hydrogen~\cite{Puchalski2019b,Puchalski2019}.
Such very accurate computations have recently been performed for vibrational transitions in tritium-bearing hydrogen molecular isotopologues~\cite{Pachucki2022d}.
On the experimental side a series of measurements has been reported for the benchmark quantity of the dissociation and ionization energies of H$_2$~\cite{Cheng2018}, D$_2$~\cite{Hussels2022}, and HD~\cite{Hoelsch2023} now reaching accuracy at the 1 MHz level and in agreement with theory.

As an alternative to the measurements based on electronic excitation, studies of vibrational splittings in the (X$^1\Sigma_g^+$) electronic ground state of molecular hydrogen may lead to improved accuracy and more advanced tests of theory in view of the long lifetimes of the states involved. 
The weak dipole moment in hydrogen deuteride (HD) heteronuclear species gives access to dipole-allowed vibrational transitions, measured via Doppler-broadened absorption spectroscopy~\cite{Kassi2011,Fasci2018,Kassi2022}, and recently also in 
saturation spectroscopy~\cite{Tao2018,Cozijn2018}.
Exploitation of the full potential of the extremely narrow Lamb dips is hampered by the asymmetry in the observed line shapes. 
This was ascribed to underlying and unresolved hyperfine structures~\cite{Diouf2019}, but also to other phenomena such as the effect of standing waves in intra-cavity experiments~~\cite{Lv2022,Jozwiak2022}.

The HT isotopologue is similarly accessible for dipole transitions.
The gas loads required to operate molecular beams cannot be supported by the exempted activity limit (1 GBq, corresponding to 11.3 mbar$\cdot$cm$^{3}$ for T$_2$) allowed in any laser laboratory inside the European Union. Experience with spectroscopic measurements on small samples of tritium containing hydrogen molecules was recently gained by our collaborative team in performing Coherent Anti-Stokes Raman (CARS) measurements on static samples of T$_2$~\cite{Trivikram2018}, DT~\cite{Lai2019} and HT~\cite{Lai2020} achieving frequency accuracies of 10 MHz in vibrational splittings. 
Before, benchmark studies had been performed via spontaneous Raman~\cite{Veirs1987} and via intracavity laser absorption~\cite{Chuang1987}.

In rovibrational transitions the HT isotopologue appears to have an advantageous hyperfine structure when compared to HD. It is simpler with less components, as $I_T=1/2$ compared to $I_D=1$, but more importantly, rovibrational transitions in HT exhibit a single isolated hyperfine component that can possibly be resolved in a high resolution saturated absorption study~\cite{Jozwiak2021d}. The present experimental study exploits this advantage.

\begin{figure}[t]
\begin{center}
\includegraphics[width=0.98\linewidth]{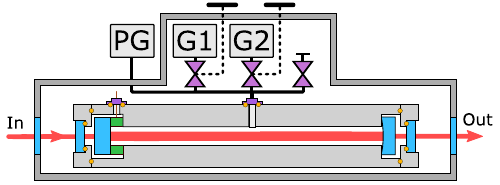}
\caption{\label{Setup}
Schematic of the tritium-adapted setup: The optical cavity with two reflective mirrors, a piezo stack and two sealed optical windows, is fully metal-sealed using indium wire seals (orange). The assembly is inside a secondary vacuum for additional stability and safety. Getter G1 is the tritium source to acquire and store the tritium. 
Getter G2 serves as an additional backing pump for when the experiment is not active. The third valve enabled initial pump-down. PG: pressure gauge.}
\end{center}
\end{figure}

Saturation spectroscopy of HT rovibrational lines is performed in an optical configuration similar to the one used for the HD experiments~\cite{Cozijn2018,Diouf2019}.
However, a major redesign (see Fig.~\ref{Setup}) was undertaken in order to operate a Noise-Immune Cavity-Enhanced Optical Heterodyne Molecular Spectroscopy (NICE-OHMS) experiment under radiation-safe conditions. 
Two layers of gas sealing windows (wedged and angled) were used for radiation safety.
A small bore diameter ($d= 10$ mm) cavity was coupled to two connected vacuum sealable chambers containing sintered porous getter materials (SAES, St171). 
These getter materials have the property that they perform as an active pump for all gases when cooling down, but do release only hydrogen when heated to temperatures in the range 600-900 $^{\rm o}$C. 
A first hydrogen-loaded getter was prepared at the TLK tritium gas-mixture infrastructure TRIHYDE~\cite{Niemes2021} with a H$_2$:HT:T$_2$ mixture (with H:T= 1:1) equivalent to 1 GBq activity. Since the getter stores hydrogen in atomic form, pure samples of HT could not be operated with~\cite{Hermann2024}. 
After loading the getter the tritium was transported to LaserLab Amsterdam and connected to the multi-chamber vacuum system. 
These measures and preparations allowed to perform measurements during a 6-week campaign, while each day evacuating the connected optical cavity to a low base pressure by a second larger-sized getter positioned in a separate cell, functioning as a pump in the setup.
Before starting the measurements tritiated gas was released into the optical cavity at a desired hydrogen pressure in the range 0.1-1.0 Pa by controlled heating of the first getter. 

For the HT frequency metrology studies a diode laser (TOPTICA DL Pro) running in the range 1430-1520 nm is used.
Signal detection of the saturation spectra is accomplished via the  NICE-OHMS technique~\cite{Ma1999,Foltynowicz2008a,Axner2014a} using sideband modulation of the carrier wave, at frequency $f_c \pm f_m$ with $f_m=405$ MHz, matching the free-spectral-range (FSR) of the cavity for generating the heterodyne NICE-OHMS signal. 
Feedback mechanisms were applied, locking the laser to the optical cavity for short-term stabilization, and locking the laser to a frequency comb for canceling long-term drift and frequency measurement ~\cite{Cozijn2018,Diouf2019}, as well as DeVoe-Brewer locking of the sidebands to the cavity~\cite{Devoe1984}. 
The optical cavity of length 0.37 m is hemispherical with one curved mirror at a radius-of-curvature of 2 m, resulting in a collimated beam at a waist diameter of 1.2 mm.
With reflectivities of $R=99.996$\% a cavity finesse of 80,000 is achieved for obtaining intracavity powers of 400 W.

Further noise reduction is accomplished via slow modulation of the cavity length (at 395 Hz) at varying peak-to-peak amplitudes of up to 100 kHz and lock-in detection and demodulation at the first derivative $(1f)$.
This $1f$ signal should result in a derivative of a dispersive Lorentzian profile~\cite{Foltynowicz2009a}, hence in a symmetric line shape, as was found for saturation spectroscopy of water~\cite{Tobias2020}.

\begin{figure}[b]
\begin{center}
\includegraphics[width=\linewidth,height=0.2\textheight]{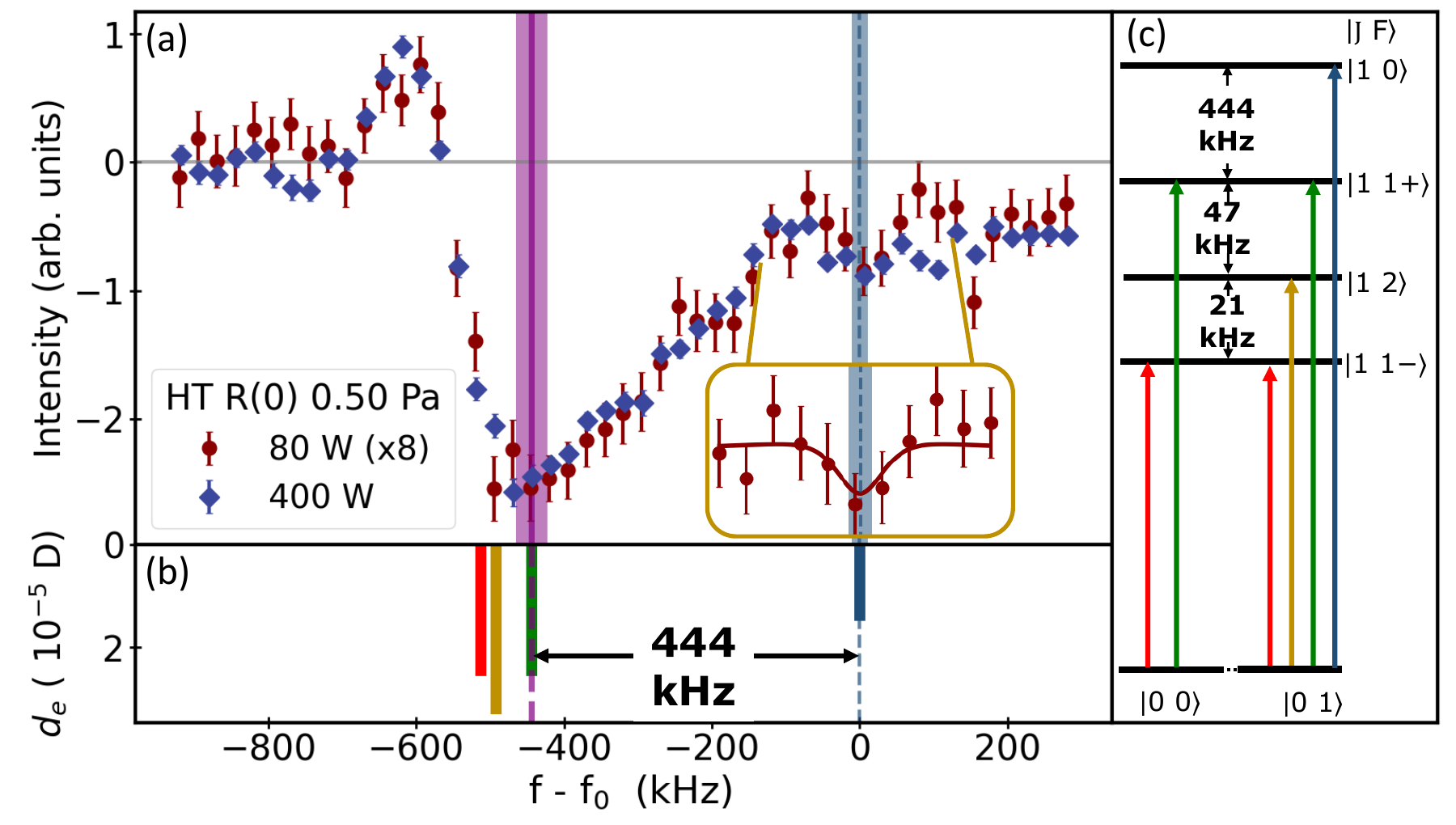}
\caption{\label{R0spec}
(a) Recorded spectra of the R(0) line at a total hydrogen pressure of 0.5 Pa (50\% HT) for intracavity powers of 80 W and 400 W. The absolute frequency scale is given via $f_0 = 203\,396\,427\,135$ kHz. In the inset the single isolated hyperfine component is shown with a fit, the width of the blue bar representing the statistical uncertainty. The width of the purple bar, at the location of the purely rovibrational transition, represents the final uncertainty. (b) Hyperfine components as a stick spectrum (strength indicated by dipole moments $d_e$); (c) Level scheme of hyperfine components.}
\end{center}
\end{figure}

Saturated absorption spectra were measured for two overtone rotational lines in the (2-0) band of HT. 
Typical recordings took some 12 hours of averaging. 
A first saturated absorption spectrum, recorded for the R(0) line, is presented in Fig.~\ref{R0spec}. 
The spectral structure consists of a broad dispersion-shaped resonance, reminiscent of the features observed in HD~\cite{Tao2018,Cozijn2018,Diouf2019,Hua2020,Cozijn2022}, and is associated with the overlapping features of 5 hyperfine components as color-coded in panels (b) and (c) of Fig.~\ref{R0spec}.
In the computations by Jozwiak et al.~\cite{Jozwiak2021d} only the couplings between nuclear spin and rotation are included, yielding the hyperfine splittings in $J=0$ rotational levels, between $F=0$ and $F=1$ components, to be zero, while in refined calculations of the spin-spin interaction a splitting of 0.3 kHz is found for HT~\cite{Puchalski2018b}. 
Unlike the case of HD there is a single isolated hyperfine component, for excitation to the $|10>$ level, located at the positive side of the hyperfineless position by 444 kHz~\cite{Jozwiak2021d}.
This component, 
observed as a symmetric narrow Lamb-dip in the spectrum measured at the lower power of 80 W, is fitted (see inset in Fig.~\ref{R0spec}; only the data points covering the isolated hyperfine component as displayed in the insets are included in the fit) to the expected line shape of a $1f$-demodulated NICE-OHMS signal~\cite{Foltynowicz2009a,Cozijn2023}:
\begin{equation}
    F(f)_{1f} = \frac{4\,A\left[\Gamma^2 - 4(f - f_0)^2 \right]}{\left[ \Gamma^2 + 4 (f - f_0)^2 \right]^2},
    \label{Eq-NO_1f}
\end{equation}
delivering the line position $f_0$ of the isolated resonance and its width  $\Gamma$.

\begin{figure}[b]
\begin{center}
\includegraphics[width=\linewidth,height=0.2\textheight]{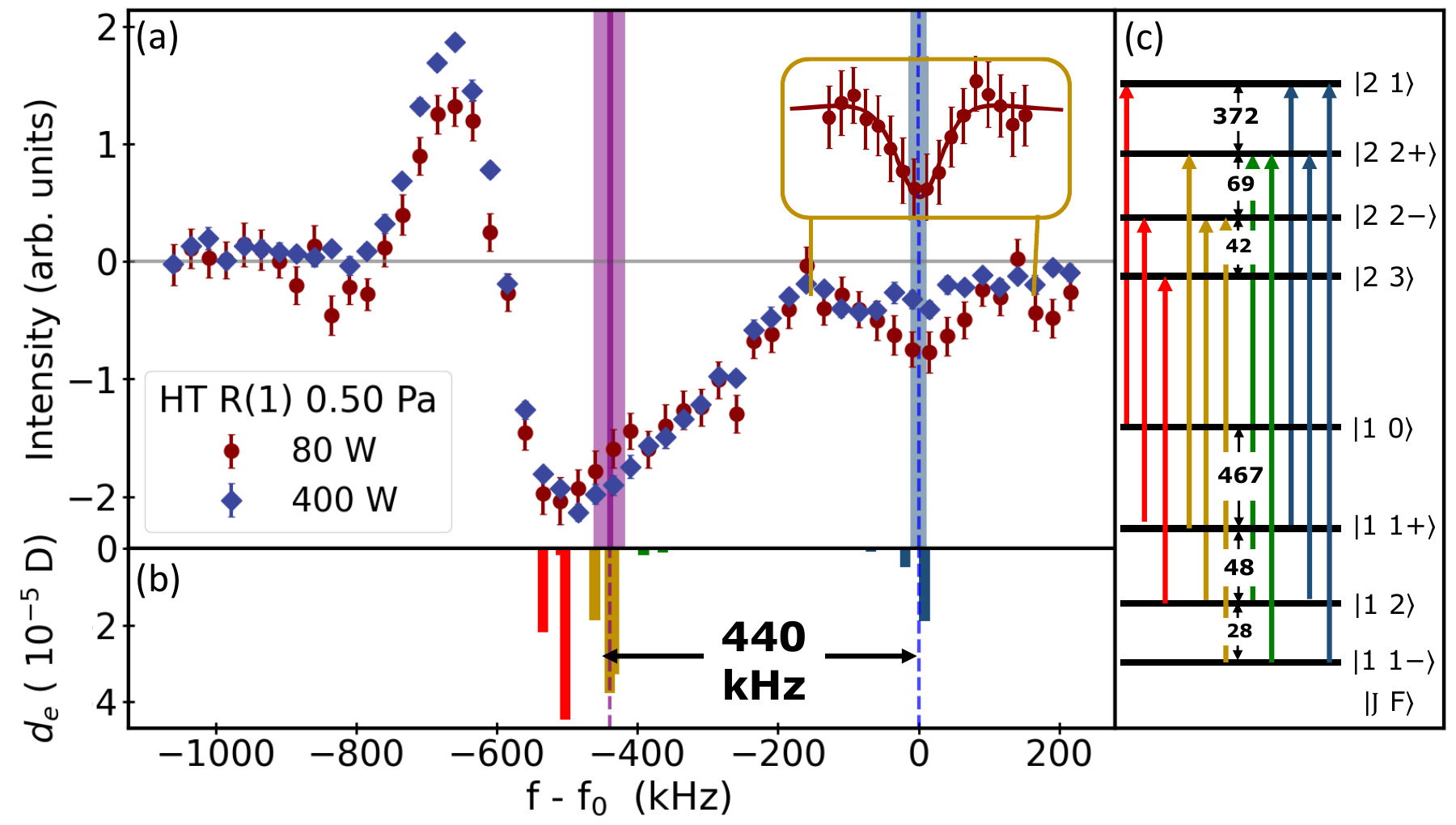}
\caption{\label{R1spec} (a) Recorded spectra of the R(1) line at two different powers, with an isolated Lamb dip at the high frequency side covering three hyperfine components. The absolute frequency scale is given via $f_0 = 205\,380\,034\,083$ kHz, corresponding to the fitted center position of the isolated Lamb dip as shown in the inset. Its width (blue bar) represents the sum of the statistical uncertainty and the uncertainty contribution of the three components (see text). The purple bar represents the location and final uncertainty of the purely rovibrational transition frequency. (b) Hyperfine stick spectrum for this transition. (c) Level scheme of hyperfine components; the numbers in between levels denote the level splittings in units of kHz. 
}
\end{center}
\end{figure}

A second spectrum, of the R(1) line in the (2-0) band, is displayed in Fig.~\ref{R1spec}. 
Again, at the high-frequency side of a broad dispersion-shaped feature, an isolated Lamb dip component appears, specifically in the low-power recording (at 80 Watt intracavity). Its center position was determined from a fit to Eq.~(\ref{Eq-NO_1f}), again for the data displayed in the inset.
For the R(1) line that component corresponds not to just one but to three hyperfine components. 

\begin{table*}
\caption{\label{Tab:Results}
Results of the experimental transition frequencies of the purely rovibrational (hyperfineless) lines R(0) and R(1) in the (2-0) band of HT, and a comparison with results  from a calculation in which the non-relativistic energy is computed  via the direct non-adiabatic approach~\cite{Pachucki2022d}. 
Differences are also presented in terms of the combined standard deviation ($\sigma$) of experiment and theory. 
}
\begin{tabular}{lcccc}
\hline
 Line & Exp. (kHz) &   H2SPECTRE~\cite{SPECTRE2022} (MHz) &   Diff. (MHz) &   Diff. ($\sigma$) \\
\hline
R(0) &  203 396 426 692 (21)  & 203 396 424.9 (9) & 1.8 (1.1) & 1.8 \\
R(1) &  205 380 033 644 (22)  & 205 380 031.7 (9) & 1.8 (1.0) & 1.8  \\
P(1) &  198 824 820 600 (100)  & 198 824 819.0 (9) & 1.9 (1.1) & 1.7 \\
\hline
\end{tabular}
\end{table*}

The spectroscopic analysis of the HT overtone resonances depend on their favorable ordering of hyperfine components, exhibiting an isolated narrow feature, not found in HD~\cite{Cozijn2018,Diouf2019}, where all hyperfine components are overlapped.
The statistical uncertainty of the fit of the R(0) line amounts to 11 kHz, while that of the R(1) is 6 kHz, due to its better signal-to-noise ratio.
No sign of asymmetry is found in the fitting procedures to the experimental data, strengthening the assumption that the isolated components represent absorption Lamb dips.
The resulting values are transformed into frequencies of the hyperfineless or purely rovibrational transitions by including the shift of the isolated hyperfine component based on accurately computed values~\cite{Jozwiak2021d} as discussed above and illustrated in Figs.~\ref{R0spec} and \ref{R1spec}.
Since for the R(1) line the isolated Lamb dip is composed of 3 overlapping hyperfine components a weighted average is taken, adding an uncertainty of 6 kHz.
The saturated absorption spectra do not exhibit a first order Doppler effect, but are subject to a second order (relativistic) Doppler effect of $f_{\rm 2D}=h\nu_0^2v^2/2mc^2$, which amounts to $1.4$ kHz for HT at room temperature.

The overall pattern of the isolated components in the spectra show that the spectra overlap for the two pressure recordings.
Quantitative analysis reveals that a pressure shift should be on the order of 5 kHz, and is conservatively estimated at 10 kHz, which is in agreement to pressure shifts found in HD (at 10 kHz/Pa)~\cite{Cozijn2018,Diouf2019} and in H$_2$ (at 15 kHz/Pa)~\cite{Cozijn2023}. 
Considering these systematic effects and contributions to the error budget final transition frequencies and uncertainties are determined
and listed in Table~\ref{Tab:Results}.

\begin{figure}[t]
\begin{center}
\includegraphics[width=\linewidth,height=0.2\textheight]{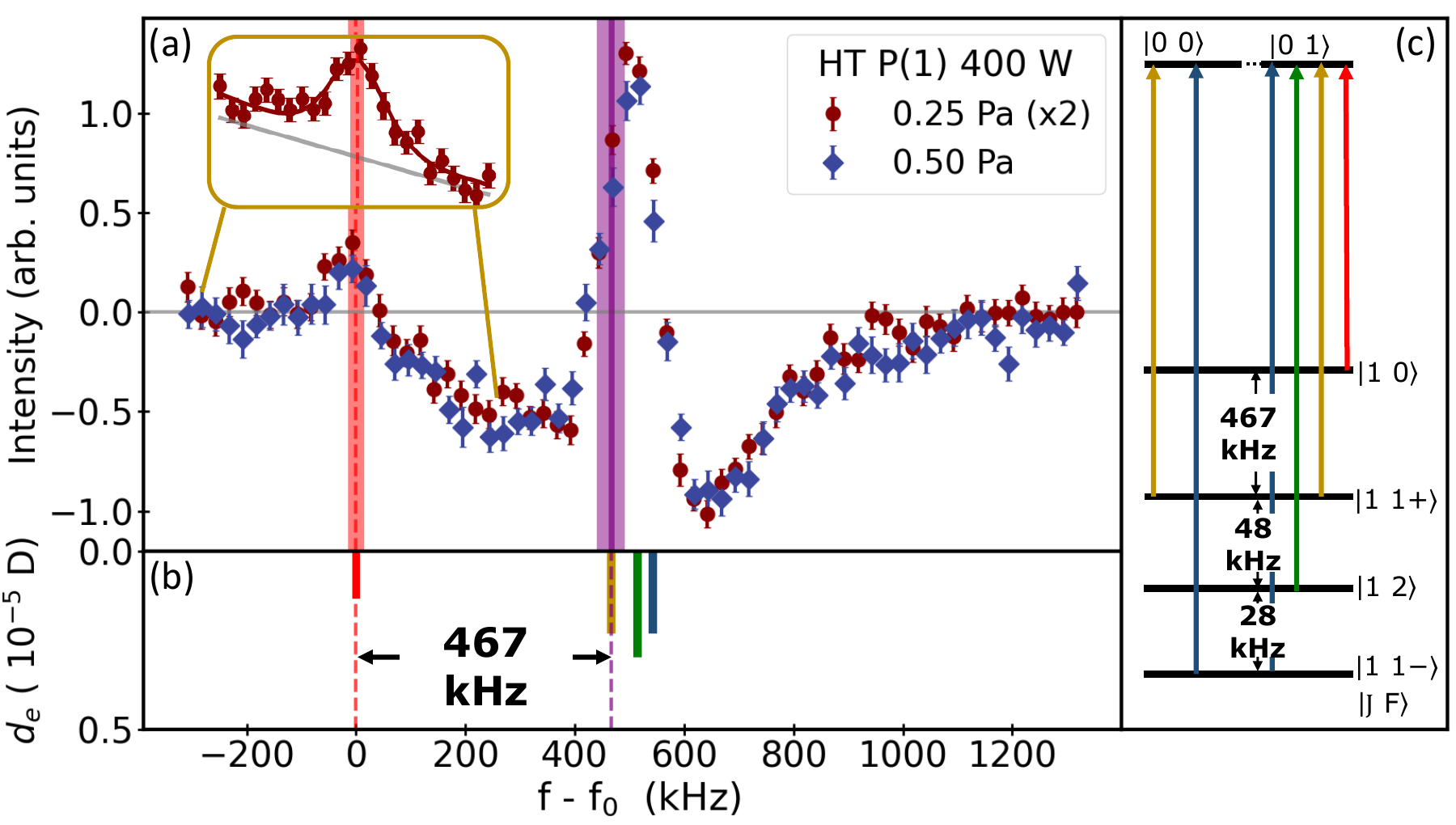}
\caption{\label{P1spec} (a) Recorded saturated absorption spectra of the P(1) line in the (2-0) band of HT for an intracavity circulating power of 400 W and hydrogen pressures as indicated.
The absolute frequency scale is given via $f_0 = 198\,824\,820\,185$ kHz. 
(b) Stick spectrum of hyperfine components.
(c) Level scheme of hyperfine components.
}
\end{center}
\end{figure}

A spectrum of the P(1) line in HT was recorded as well at conditions as indicated in Fig.~\ref{P1spec}. Similar to the case of the P(1) line in HD~\cite{Diouf2020} the saturated spectral components appear as Lamb peaks, rather than as Lamb dips. 
In the case of HD this phenomenon of reversed signal amplitudes was interpreted as originating from the contribution of crossover resonances, that were included in a comprehensive model based on optical Bloch equations~\cite{Diouf2019}. 
An important learning from the present study is that this cannot be the case for HT, as crossovers cannot interfere with regular hyperfine components at the location of the isolated component. 
It must be concluded that the present finding contradicts an explanation based on crossover resonances for the signal inversion in HD~\cite{Diouf2020}.
We do not have an explanation for the phenomenon of signal inversion, neither in HT nor in HD, which adds to the conundrum of the dispersive-like line shapes of the strong saturation signals of the non-isolated hyperfine components in both isotopologues. 
Future studies may reveal whether these phenomena are related to 
optical pumping, collisional relaxation, or trapping effects of molecules in the nodes of intense standing waves in the cavity.
In view of the fact that P(1) is not observed as a regular Lamb dip, we do not pursue extracting an accurate transition frequency via fitting. 
A rovibrational transition frequency is preliminary estimated at $198\,824\,820\,600$ kHz with a conservative uncertainty estimate of 100 kHz.

A note must be made on the recoil effect on the measured transitions. As is known from the early days of laser spectroscopy in the saturation regime both a red and a blue shifted recoil component is observed~\cite{Hall1976b,Barger1979,Bagayev1991} at frequencies $f_{\rm rec} =  \pm h\nu_0^2/2mc^2$.
For the case of the R-lines of HT in the (2-0) band the recoil shift corresponds to $\pm 22$ kHz. 
Recently, in the case of a measurement of a Lamb dip in the very weak quadrupole spectrum of H$_2$~\cite{Cozijn2023} only a single recoil component was observed, which was interpreted as the blue recoil. 
In the present case of HT the combination of widths and noise levels do not allow for a decision whether the observed spectra cover both recoil components or only one.
Here we follow the generally accepted approach to the recoil phenomenon in saturation that the blue and red recoil components average out to no shift of the molecular resonance frequency.
If the suppression of red recoil components in intracavity saturation spectroscopy of weak transitions is an ubiquitous feature, then the measured frequencies (as in Table~\ref{Tab:Results}), should be lowered by 22 kHz. For the moment this is left as an open issue.

The present experimental values for the transition frequencies in the (2-0) band come timely for a comparison with very recent calculations for the tritium bearing hydrogen isotopologues~\cite{Pachucki2022d}. 
A recent calculation for based on a direct non-adiabatic approach (DNA, hence a four-particle variational~\cite{Puchalski2019,Puchalski2019b} 
involving nonadiabatic James-Coolidge wave functions led to an improvement by an order of magnitude, reaching 0.9 MHz for the overtone transitions in HT~\cite{Pachucki2022d}.
The DNA method was employed to compute non-relativistic energies at an accuracy of 10$^{-13}$, so exact for the purpose of comparison with present state-of-the art experiments. 
The results from the DNA non-relativistic approach are augmented with computations of relativistic and quantum electrodynamical (QED) corrections~\cite{Komasa2019}, which produce the final theoretical uncertainties. 

These data have been incorporated in the H2SPECTRE program, version 7.4~\cite{SPECTRE2022}.
Results of the H2SPECTRE suit indicate that the uncertainty in the computation of level energies amounts to $1.4 \times 10^{-4}$ cm$^{-1}$ (or 4.2 MHz), but cancelation of uncertainty occurs when computing the transition frequencies yielding uncertainties of $3.1 \times 10^{-5}$ cm$^{-1}$ (or 0.9 MHz).
H2SPECTRE also reveals the origin of the estimated uncertainty, which is in the $E^{(5)}$ or $m\alpha^5$-term in the level energy expansion of the fine structure constant, the leading order QED correction.
In Table~\ref{Tab:Results} results from experiment and theory are compared for the lines measured in the present study.

The data in  Table~\ref{Tab:Results} show that the accuracy of the present experimental transition frequencies is among the highest in the hydrogen isotopologues, with those of HD~\cite{Fast2020} and H$_2$~\cite{Cozijn2023}.
Again, a systematic deviation is found between experiment and theory at the level of $1.8\sigma$.
Previously, deviations at exactly the same $1.8\sigma$ were found for the (2-0) vibrational bands in HD~\cite{Fast2020,Cozijn2022b}, in H$_2$~\cite{Fleurbaey2023,Cozijn2023}, and in D$_2$~\cite{Zaborowski2020}. 
This large set of data makes the deviations significant.
Even for the more advanced treatment of the leading order QED term, that was recently recomputed~\cite{Silkowski2023}, the discrepancies persist.

The present study exploits the favorable hyperfine structure in the tritium hydride species using an isolated hyperfine component in the R(0) and R(1) transitions in the (2-0) band to determine purely rovibrational transition frequencies at an accuracy of 20 kHz, a thousand times better than for any previously observed transition in a tritiated hydrogen molecule.
It is shown that measurement campaigns of extended duration can be performed in a cavity-enhanced configuration without radioactive tritium degrading the highly reflective mirror coatings, nor having detrimental effects on the instrumentation. This opens the perspective of adding all three tritiated hydrogen molecules to the set of benchmark systems for testing molecular quantum theory and probing new physics~\cite{Ubachs2016}.
With the recent computations of non-relativistic energies in the tritiated species~\cite{Pachucki2022d}, the current results of HT frequencies form an accurate test ground for future computations of the $m\alpha^5$ nonadiabatic QED term in molecular hydrogen isotopologues.

\vspace{0.5cm}
Acknowledgment. The research was funded via the Access Program of Laserlab-Europe (Grant Numbers 654148 and 871124), a European Union’s Horizon 2020 research and innovation programme. Financial support from the Netherlands Organisation for Scientific Research (NWO), via the Program “The Mysterious Size of the Proton” is gratefully acknowledged. M. Schlösser wishes to thank the Baden-Württemberg Foundation for the generous support of this work within the Elite-Postdoc-Fellowship. We thank the KIT mechanical workshop for constructing the multiple chamber cell structure, the co-workers at TLK for supporting tritium activities (safety, legal, filling, advise), and the radiation safety officer at VUA (Ms. J. Houben-Weerts) for making this experiment possible.

\bibliography{Hydrogen,NICE-OHMS,Tritium}

\end{document}